\documentclass[fleqn,twocolumns,usenatbib]{mnras}

\usepackage[T1]{fontenc}
\usepackage{ae,aecompl}

\usepackage{graphicx}	
\usepackage{amsmath}	

\usepackage{amssymb}	
\title[Sulfur X-ray ISM absorption]{Sulfur X-ray absorption in the local ISM}

\author[Gatuzz et al.]{
Efrain Gatuzz$^{1}$\thanks{E-mail: egatuzz@mpe.mpg.de} 
T. W. Gorczyca$^{2}$,
M. F. Hasoglu$^{3}$,
E. Costantini$^{4,5}$,\newauthor
Javier~A.~Garc\'ia$^{6,7}$
and Timothy~R.~Kallman$^{7}$ 
\\
$^{1}$Max-Planck-Institut f\"ur extraterrestrische Physik, Gie{\ss}enbachstra{\ss}e 1, 85748 Garching, Germany\\
$^{2}$Department of Physics, Western Michigan University, Kalamazoo, MI 49008, USA\\
$^{3}$Department of Computer Engineering, Hasan Kalyoncu University, 27100 Sahinbey, Gaziantep, Turkey\\
$^{4}$SRON Netherlands Institute for Space Research, Niels Bohrweg 4, NL-2333 CA Leiden, the Netherlands \\
$^{5}$Anton Pannekoek Institute, University of Amsterdam, Postbus 94249, NL-1090 GE Amsterdam, the Netherlands \\
$^{6}$Cahill Center for Astronomy and Astrophysics, California Institute \\
$^{7}$NASA Goddard Space Flight Center, Greenbelt, MD 20771, USA\\
}

\date{Accepted XXX. Received YYY; in original form ZZZ}
\pubyear{2020}
\begin{document}

 \label{firstpage}
\pagerange{\pageref{firstpage}--\pageref{lastpage}}
\maketitle
\begin{abstract} 
We present a study S K-edge using high-resolution HETGS {\it Chandra} spectra of 36 low-mas X-ray binaries. For each source, we have estimated column densities for {\rm S}~{\sc i}, {\rm S}~{\sc ii}, {\rm S}~{\sc iii}, {\rm S}~{\sc xiv}, {\rm S}~{\sc xv} and {\rm S}~{\sc xvi} ionic species, which trace the neutral, warm and hot phases of the Galactic interstellar medium. We also estimated column densities for a sample of interstellar dust analogs. We measured their distribution as a function of Galactic latitude, longitude, and distances to the sources. While the cold-warm column densities tend to decrease with the Galactic latitude, we found no correlation with distances or Galactic longitude. This is the first detailed analysis of the sulfur K-edge absorption due to ISM using high-resolution X-ray spectra.
\end{abstract}


\section{Introduction}\label{sec_in}
The interstellar medium (ISM) is one of the essential components in Galactic dynamics because it regulates the star life cycles as well as cooling and Galactic star formation rates\citep{won02,big08,ler08,lad10,lil13}. Defined as gas and dust between stars, the ISM shows multiple phases characterized by different gas temperatures, which vary from 10 to 10$^{6}$ K \citep[e.g.][]{mck77,fal05,ton09,dra11,jen11,rup13,zhu16,sta18}. 

Such a complex environment can be analyzed using the high-resolution X-ray spectroscopy technique. Bright X-ray sources, acting as lamps, are required to carry out such an analysis. The physical properties of the gas between the observer and the source are studied by modeling the absorption features identified in the X-ray spectra. In the last decade, the O, Fe, Ne, Mg, N, and Si K absorption edges associated with the ISM have been analyzed by applying this method  \citep{pin10,pin13,cos12,gat13a,gat13b,gat14,gat15,gat16,joa16,gat18a,gat18b,gat18c,gat19,zee19,gat20,psa20,gat21,rog21,yan22,gat23}.

Sulfur constitutes an excellent diagnostic tool for studying various astrophysical environments, although its chemistry still needs to be fully understood \citep{laa19}. Sulfur remains in ionized atomic form within primitive ISM environments \citep{sav96b,jen09}. Sulfur-bearing molecules probe the physical structure of star-forming regions \citep{lad91,plu97}. Depletion into dust grains has been proposed due to the drastic reduction of the cosmic sulfur abundance in molecular clouds \citep{kel02,sca03,wak04}. 

Here we present an S K-edge absorption region analysis using {\it Chandra} observation of low-mass X-ray binaries (LMXBs). Section~\ref{sec_xray_data} describes the data sample and spectral fitting procedure. Section~\ref{sec_s_atom} describes the atomic data calculation and photoabsorption cross-sections included in the modeling. Section~\ref{sec_dis} discusses the results obtained from the fits. Finally, we summarize the main results of our analysis in Section~\ref{sec_con}.

\begin{figure} 
\centering
\includegraphics[scale=0.31]{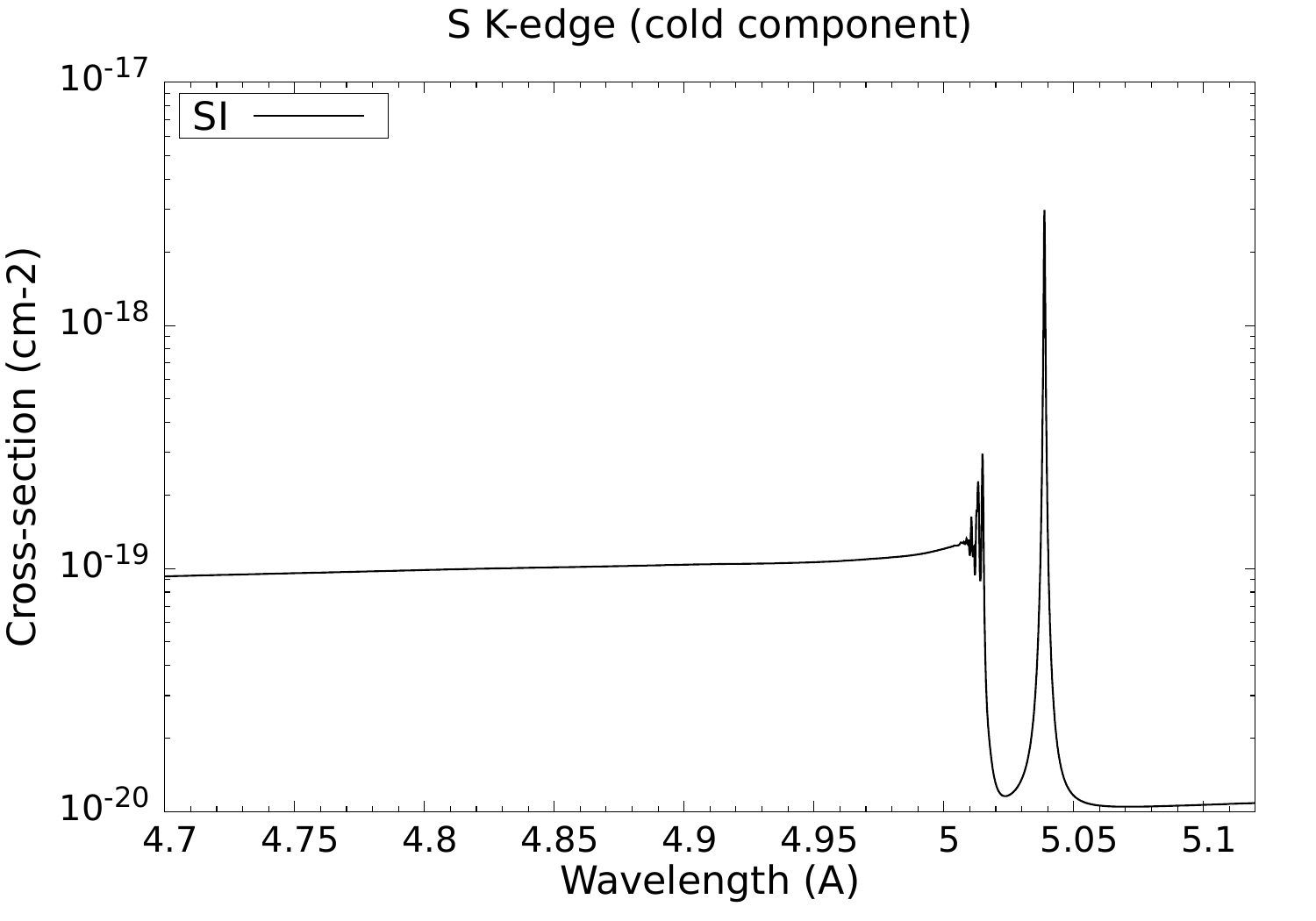}\\
\includegraphics[scale=0.31]{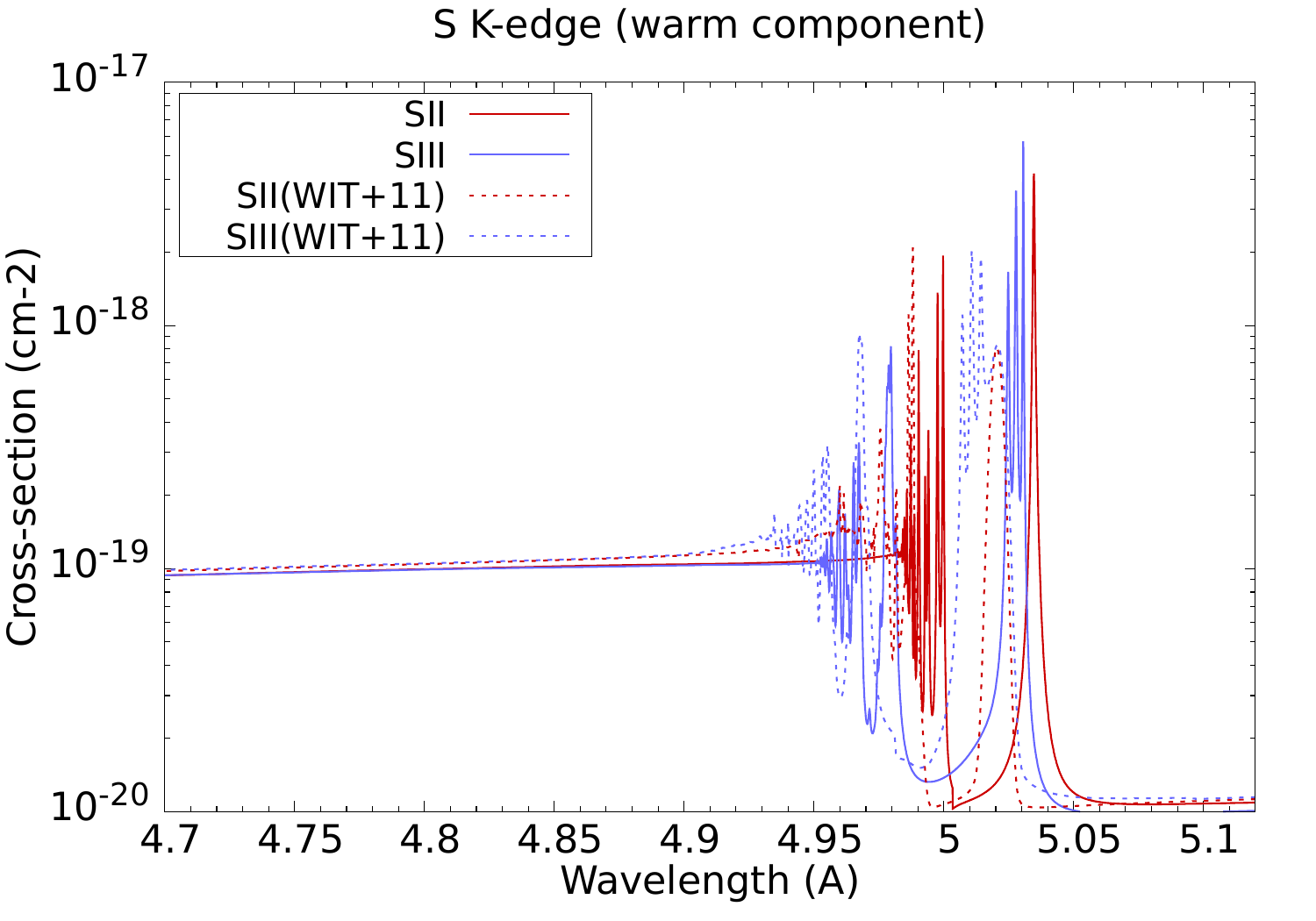}\\
\includegraphics[scale=0.31]{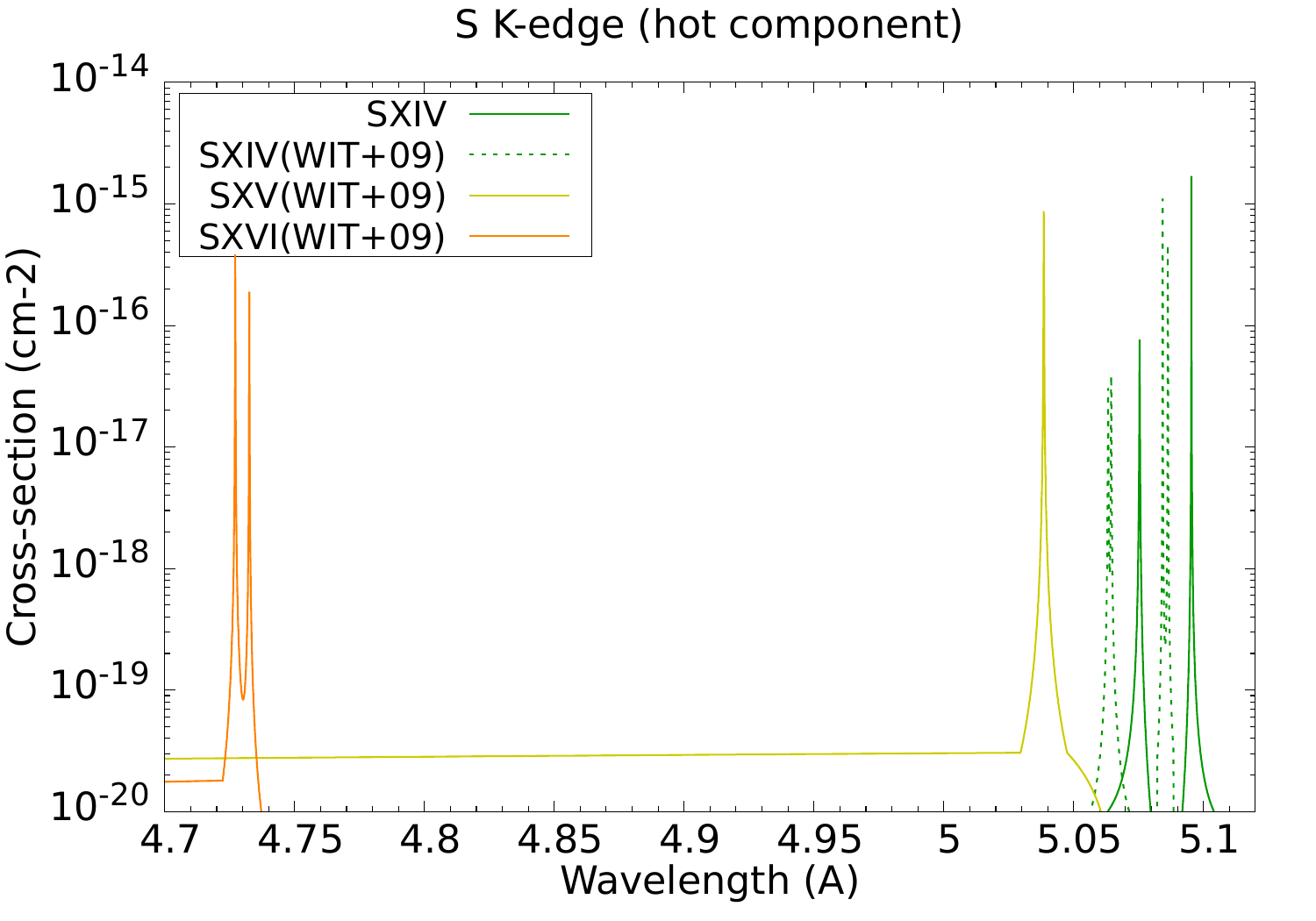}\\
\includegraphics[scale=0.31]{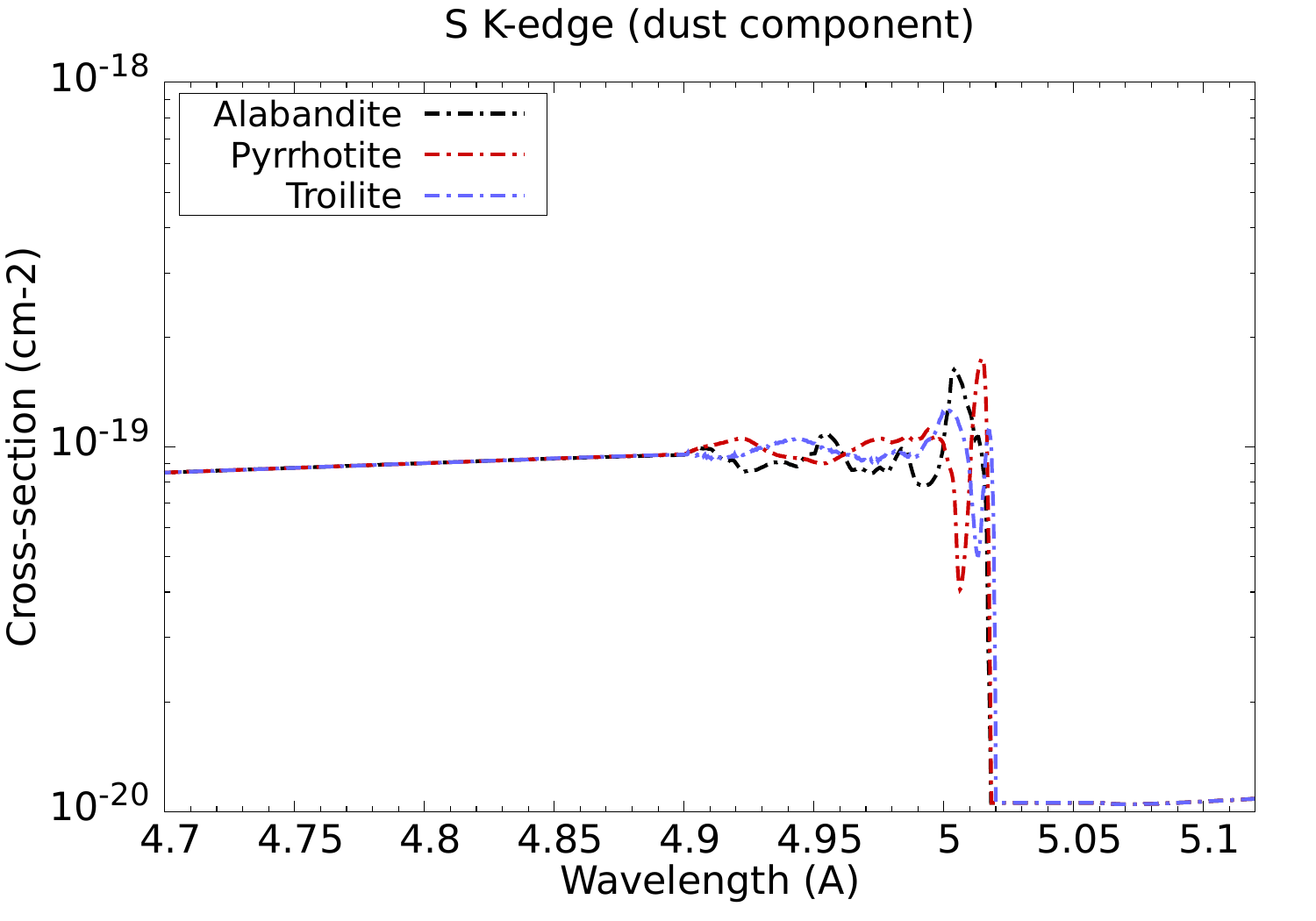}
\caption{Photoabsorption cross sections included in the model for {\rm S}~{\sc i} (top panel), {\rm S}~{\sc ii}, {\rm S}~{\sc iii} (second panel), {\rm S}~{\sc xiv}, {\rm S}~{\sc xv} and {\rm S}~{\sc xvi} (third panel) and three samples of interstellar dust analogs, namely alabandite, pyrrhotite and troilite (bottom panel). Photoabsorption cross-sections, previously computed by \citet{wit09,wit11}, are also shown. While the general profiles are quite similar, differences exist in the K-edge positions between the two calculations because orbital relaxation effects are considered in the present work for generating a basis tailored to inner-shell ionization.}\label{fig_s_cross}
\end{figure}

.

\section{X-ray observations and spectral fitting}\label{sec_xray_data} 
We analyze {\it Chandra} spectra of 36 low-mass X-ray binaries (LMXBs) along different lines of sight. To build our sample, we have selected those sources for which we have at least 1000 counts in the sulfur edge absorption region (4.5--5.5 \AA). To get an unbiased sample, we did not impose any constraints on the significance of the detection for a particular line (e.g. {\rm S}~{\sc xvi} K$\alpha$ detection). Table~\ref{tab_lmxbs} shows the specifications of the sources, including Galactic coordinates, hydrogen column densities taken from \citet{wil13}, and distances, if available. We note that \citet{wil13} column densities correspond to $N({\rm HI+H2})$ average values along line-of-sights covering all material in the galaxy. However, they can be safely used because our analysis covers a small wavelength region ($\sim 1$ \AA).

All spectra have been obtained with the High Energy Grating (HEG) from the high-energy transmission grating (HETGS) in combination with the Advanced CCD Imaging Spectrometer (ACIS). Observations were reduced using the Chandra Interactive Analysis of Observations (CIAO\footnote{http://cxc.harvard.edu/ciao/threads/gspec.html}, version 4.15.1), including background subtraction and following the standard procedure. All observations were fitted with the {\sc xspec} package (version 12.11.1\footnote{http://heasarc.nasa.gov/xanadu/xspec/}) in the 4.5--5.5 \AA\ wavelength region. We modeled the continuum with a {\tt powerlaw} model. For each source, all observations were fitted simultaneously with $\gamma$ and the normalization as independent parameters to account for changes in the continuum at different epochs. Finally, we use the $\chi^{2}$ statistics in combination with the \citet{chu96} weighting method.

\section{Atomic data calculation}\label{sec_s_atom} 
To fit the S K-edge absorption region, we computed {\rm S}~{\sc i}-{\rm S}~{\sc xiv} photoabsorption cross-sections (i.e., Li-like) as follows. From a single-configuration perspective for the inner-shell photoexcitation of the sulfur ground state, the specific processes to be considered are the followings:
\begin{eqnarray}
h\nu+S(1s^22s^22p^63s^23p^4)[ ^3P ]\rightarrow 1s2s^22p^63s^23p^4np[ ^3S^o, ^3P^o, ^3D^o ]\ .
\end{eqnarray}
This intermediate autoionizing, or resonant, state can decay via two qualitatively different Auger pathways. Firstly, there is {\em participator} Auger decay
\begin{eqnarray}
1s2\ell^{\, 8} 3\ell^{\, 6} np & \rightarrow & 1s^2 2\ell^{\, a} 3\ell^{\, b} +e^-\ , \ \ (a+b=13) \nonumber \ ,
\label{eqpart}
\end{eqnarray}
in which the valence electron $np$ participates in the autoionization process, thus giving a decay rate that scales as $1/n^3$. On the other hand, secondly {\em spectator} Auger decay 
\begin{eqnarray}
1s2\ell^{\, 8} 3\ell^{\, 6} np & \rightarrow & 1s^2 2\ell^{\, c} 3\ell^{\, d}np +e^-\ , \ \ (c+d=12) \nonumber \ ,
\label{eqspect}
\end{eqnarray}
proceeds via a stronger, $n$-independent Auger rate, causing a massive broadening of the entire Rydberg series of resonances below the K-edge. Participator Auger decay is accounted for straightforwardly by explicitly including all final {\rm S}~{\sc ii} ionic channels in the standard R-matrix implementation \citep{burke,berrington95}.

A point needs to be made regarding an alternate, and sometimes significant, decay pathway: that of spontaneous core radiative decay.  This would occur most strongly, following the above initial photoabsorption, as the alternate $1s2s^22p^63s^23p^4np \rightarrow 1s^22s^22p^63s^23p^3np + h\nu$ decay.  However, using the structure and collision code AUTOSTRUCTURE~\citep{bad86,bad97}, it has been determined that the fluorescence branching ratio is small, less than 1\%, and any minor correction to these Auger Lorentzian widths is eventually washed out once a broader, less-certain x-ray spectral (Gaussian) resolution is considered.  Furthermore, even though there exist other R-matrix and/or AUTOSTRUCTURE calculations for the positions and widths of the associated resonances involved, it is not clear how useful a detailed comparison of these would be since there are inherent uncertainties in energy positions due to the variational principle for approximate wavefunctions.  And the more important quantity is really the integrated oscillator strength (or cross section, which tends to give much more accurate Maxwellian (say) rate coefficients as needed in astrophysical plasma modeling.

Present calculations utilize the modified $R$-matrix method \citep{berrington95,burke} to account for the spectator Auger broadening via an optical potential described by \citet{Gorczyca99}. This enhanced $R$-matrix method with pseudo resonance elimination \citet{pseudo} has been used in describing experimental synchrotron measurements for argon \citep{Gorczyca99}, oxygen \citep{goro,gor13}, neon \citep{gorne}, and carbon \citep{hasoglu_c} accurately.

The employed orbital basis consists of physical and pseudoorbitals to account for relaxation effects following $1s$-vacancy. For the system with number of electrons in target states $ N \ge 10 $, $1s$, $2s$, $2p$, $3s$, and $3p$ and $\overline{3d}$, $\overline{4s}$, and $\overline{4p}$ are treated as physical orbitals and pseudoorbitals, respectively. For the cases $ N < 10 $, using only $1s$, $2s$, and $2p$ physical orbitals and $\overline{3s}$, $\overline{3p}$, and $\overline{3d}$ pseudoorbitals are found to be sufficient. The physical orbitals are formed by using the Hartree-Fock method, and pseudoorbitals are optimized by using the multi-configuration Hartree-Fock method on the configuration lists, including single and double promotions from the $1s$-vacancy target state to account for important orbital relaxation effects due to the K-shell vacancy.

To compute Auger widths used for the spectator Auger broadening effects in $1sn\ell^q$ autoionizing states target states by using the R-matrix method, we rely on Wigner Time Delay Method \citep{smith},  as it was applied in the recent photoabsorption works on C-, Mg-, and Si-isonuclear sequences\citep{hasoglu_c,mg,gatuzz_si}. R-matrix method on $e^-+1s^2n\ell^{q-2}$ scattering problem is employed in the same manner as photoabsorption calculations in terms of basis set and configuration lists. Further details on the significance of spectator Auger broadening effects and application of R-matrix along with the Wigner Time Delay method can be found in the previous works \citep{Gorczyca99, goro, gorne, hasoglu_c,gor13,mg}.

\subsection{S~K-edge photoabsorption cross-sections}\label{sec_s_cross} 
We consider the {\rm S}~{\sc i}, {\rm S}~{\sc ii}, {\rm S}~{\sc iii} and {\rm S}~{\sc xiv} photoabsorption cross-sections computed as described above in the model. Previously reported K-absorption ($1s \rightarrow np$) cross sections of Sulfur ionized species for $N < 11$ and $N \ge 11$ were computed by \citet{wit09} and \citet{wit11}, respectively, by utilizing a similar R-matrix approach with inclusion of Auger broadening effects. However, important orbital relaxation effects were neglected due to the fact that the single-electron orbitals were obtained by using a Thomas-Fermi-Dirac statistical model potential. Inclusion of relaxation effects primarily becomes important due to the change in the potential perceived by the outer electrons upon an excitation or ionization of an electron from the K-shell; this change in potential strength becomes more significant for low-charged ions. In particular, {\rm S}~{\sc ii}  is expected to be the most affected by relaxation effects, as evidenced by the K-shell threshold being overestimated by  approximately $7$~eV (see Figure 1). Previous calculations for $N < 11$ by \citet{wit09} were performed using the Breit-Pauli R-matrix method that showed significant fine-structure splitting of resonance series (see Figure 1 for {\rm S}~{\sc xiv}), which are included here. However, such relatively small shifts in energy are essentially washed out of any convolution as needed for plasma modeling. 

We have also included extinction cross-sections (i.e., absorption$+$scattering) for three samples of interstellar dust analogs, namely alabandite, pyrrhotite, and troilite, measured by \citet{cos19b}. Figure~\ref{fig_s_cross} shows the sulfur photoabsorption cross-sections considered in the model, which takes into account the cold (top panel), warm (second panel), hot (third panel), and dust (bottom panel) phases of the ISM. Photoabsorption cross-sections computed by \citet{wit09,wit11} are also included.

\begin{figure}
          \centering  
\includegraphics[width=0.48\textwidth]{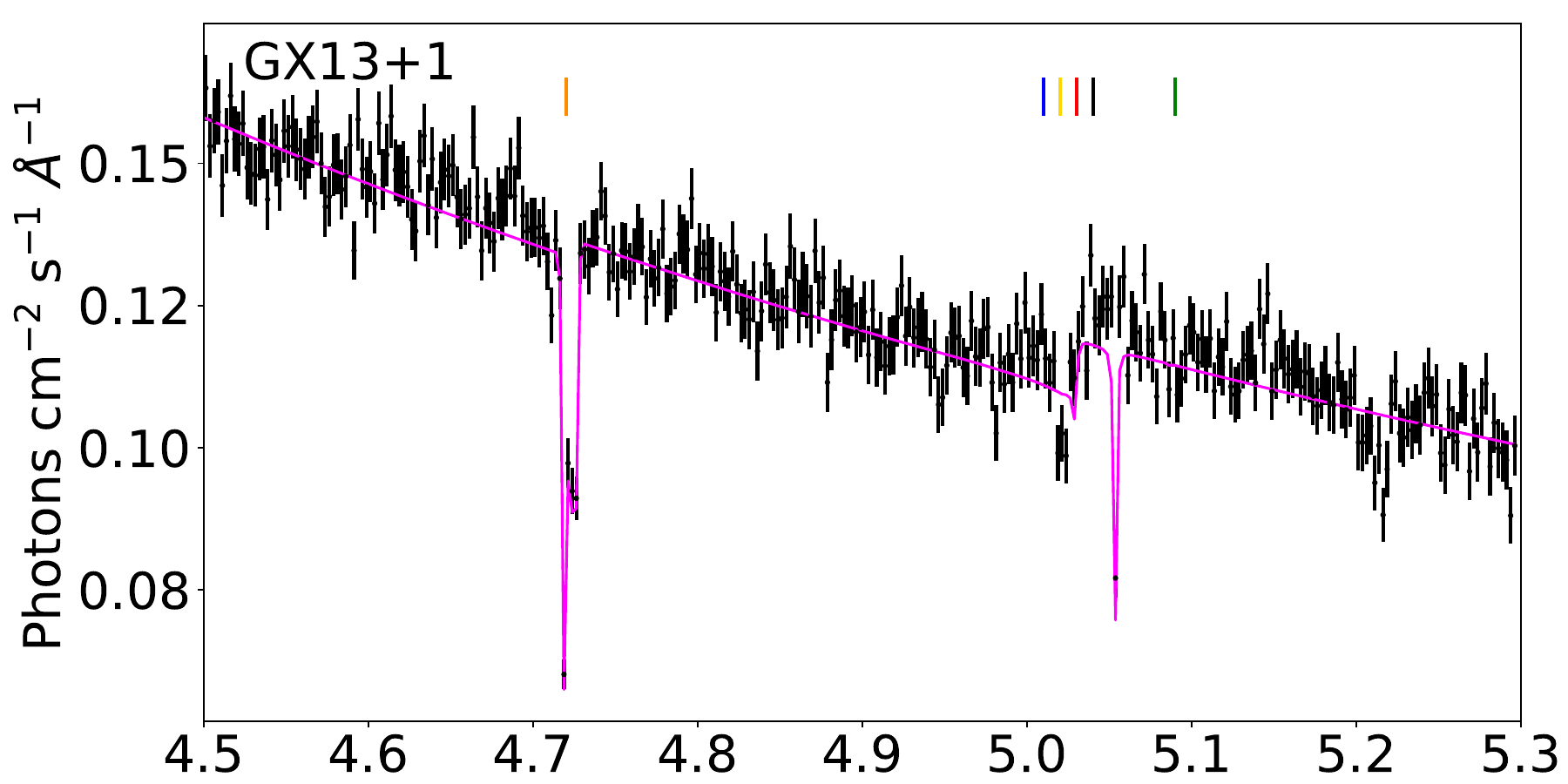} \\  
\includegraphics[width=0.48\textwidth]{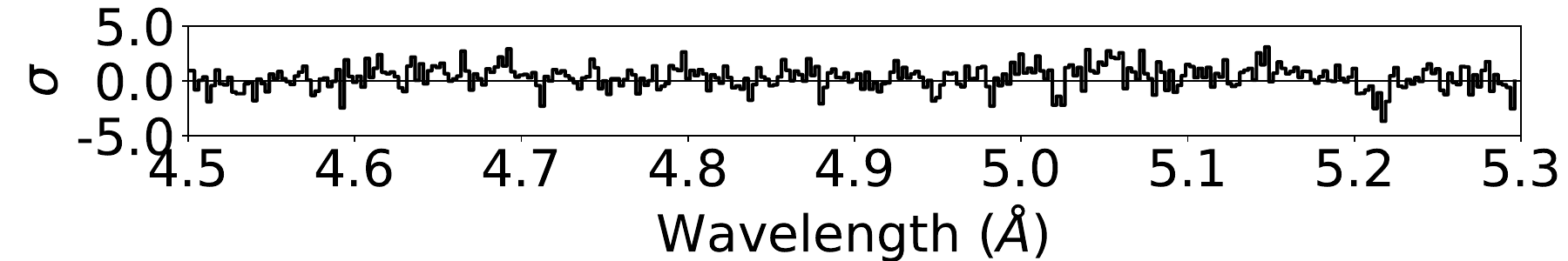}   
      \caption{
      Best-fit results in the S K-edge photoabsorption region for the LMXB GX13+1. The black data points correspond to the observations (in flux units and combined for illustrative purposes), while the solid red line corresponds to the best-fit model.  The position of the K$\alpha$ absorption lines are indicated for each ion, following the color code used in Figure~\ref{fig_s_cross}.
      }\label{fig_fits_gx13+1}
   \end{figure}

We included these cross-sections in a modified version of the {\tt ISMabs} model \citep{gat15} to model the S K-edge. In this way, the column densities for the ionic species of interest are free parameters in the spectral fitting. For each source, we fixed the ${\rm HI}$ {\tt ISMabs} column densities to the values provided by \citet{wil13}. Given the spectral resolution of the instrument, we note that a detailed benchmarking of the doublet/triplet resonance line positions (see Figure~\ref{fig_s_cross}) cannot be performed. For example, for the {\rm S}~{\sc iii} K$\alpha$ resonance lines, we have a separation of $\Delta\lambda\sim 5$ m\AA\ while the nominal HEG resolution is $\Delta\lambda\sim 12$ m\AA. For example, Figure~\ref{fig_fits_gx13+1} shows the best fit obtained for the LMXB GX13+1. Black points correspond to the observation in flux units, while the red line corresponds to the best-fit model. Residuals are included in units of $(data-model)/error$. The position of the K$\alpha$ absorption lines for the gaseous component is indicated for each ion, following the color code used in Figure~\ref{fig_s_cross}.

           \begin{figure}
          \centering
\includegraphics[width=0.43\textwidth]{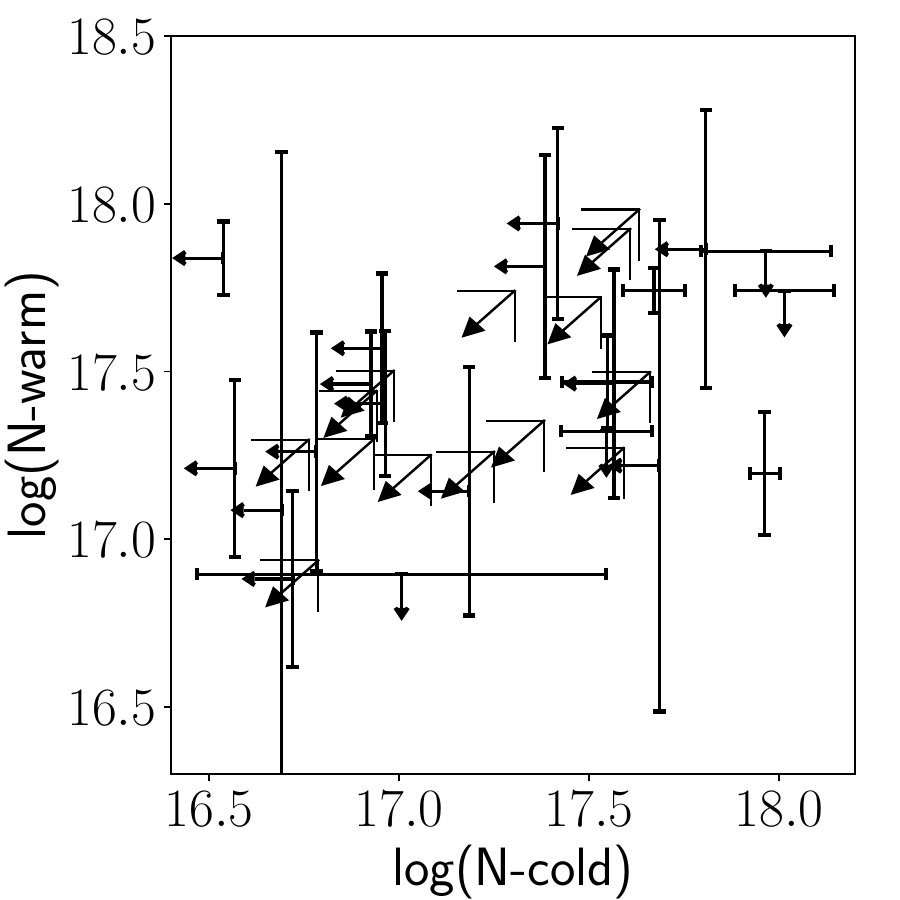}\\
\includegraphics[width=0.43\textwidth]{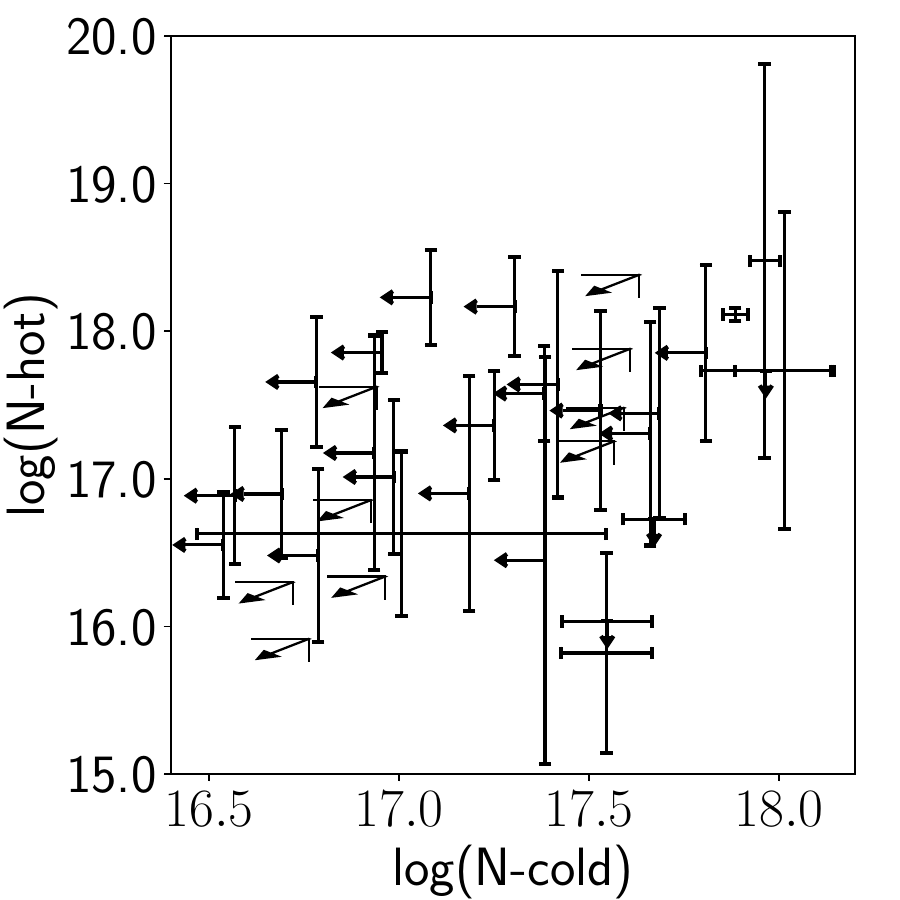}\\
\includegraphics[width=0.43\textwidth]{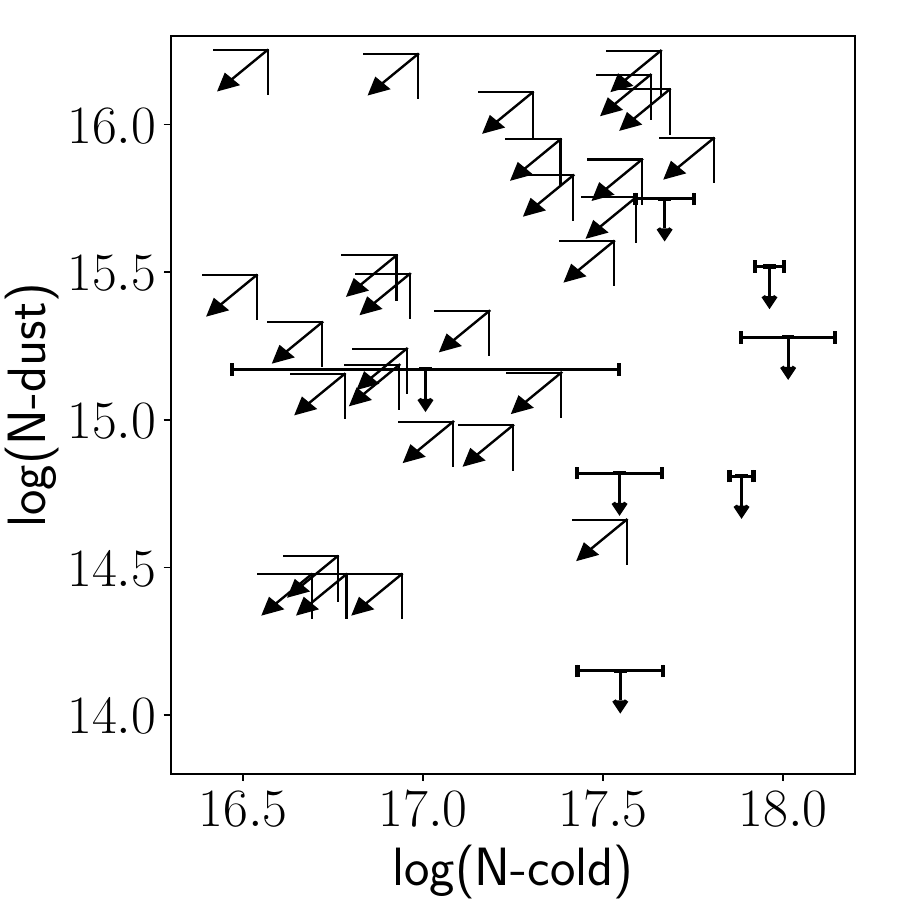}   
      \caption{
      Best fit column densities for the cold ({\rm S}~{\sc i}), warm ({\rm S}~{\sc ii}+{\rm S}~{\sc iii}), hot ({\rm S}~{\sc xiv}+{\rm S}~{\sc xv}+{\rm S}~{\sc xvi}) and dust ISM phases.
      }\label{fig_data_density_fractions}
   \end{figure}

\begin{table}
\caption{\label{tab_lmxbs}List of Galactic observations analyzed.}
\centering
\begin{tabular}{lcccccccc}
\hline
Source   & Galactic &Distance     &$N({\rm HI+H2})$\\
 &Coordinates &(kpc)   &   & \\
\hline  
4U~0614+091 &$( 200.88 , -3.36 )$&$ 2.2 _{ -0.7 }^{ +0.8 }$ $^{a}$&$ 5.86 $\\ 
4U~1254--690 &$( 303.48 , -6.42 )$&$ 13 \pm 3$ $^{b}$&$ 3.46 $\\ 
4U~1630--472 &$( 336.91 , 0.25 )$&$ 4 $ $^{c}$&$ 17.6 $\\ 
4U~1636--53 &$( 332.92 , -4.82 )$&$ 6 \pm 0.5$ $^{d}$&$ 4.04 $\\ 
4U~1702--429 &$( 343.89 , -1.32 )$&$ 6.2 \pm 0.9$ $^{e}$&$ 12.3 $\\ 
4U~1705--44 &$( 343.32 , -2.34 )$&$ 8.4 \pm 1.2$ $^{e}$&$ 8.37 $\\ 
4U~1728--16 &$( 8.51 , 9.04 )$&$ 4.4$ $^{c}$&$ 3.31 $\\ 
4U~1728--34 &$( 354.30 , -0.15 )$&$ 5.3\pm 0.8$ $^{e}$&$ 13.9 $\\ 
GX~9+9 &$( 8.51 , 9.04 )$&$ 4.4 $ $^{c}$&$ 3.31 $\\ 
H1743--322 &$( 357.26 , -1.83 )$&$ 10.4 \pm 2.9$ $^{f}$&$ 8.31 $\\ 
IGRJ17091--3624 &$( 349.52 , 2.21 )$&$ -$&$ 7.72 $\\ %
NGC~6624 &$( 2.79 , -7.91 )$&$ 7 $ $^{g}$&$ 2.33 $\\ 
EXO~1846--031 &$( 29.96 , -0.92 )$&$-$&$ 13.9 $\\ %
GRS~1758--258 &$( 4.51 , -1.36 )$&$ 8.5$ $^{c}$&$ 9.03 $\\ 
GRS~1915+105 &$( 45.37 , -0.22 )$&$ 11 _{ -4 }^{ +1 }$ $^{e}$&$ 15.1 $\\ 
GS~1826--238 &$( 9.27 , -6.09 )$&$ 7.5 \pm 0.5$ $^{h}$&$ 3.00 $\\ 
GX~13+1 &$( 13.52 , 0.11 )$&$ 7\pm 1$ $^{i}$&$ 13.6 $\\ 
GX~17+2 &$( 16.43 , 1.28 )$&$ 14 _{ -2.1 }^{ +2 }$ $^{e}$&$ 10.0 $\\ 
GX~3+1 &$( 2.29 , 0.79 )$&$ 5 _{ -0.7 }^{ +0.8 }$ $^{j}$&$ 10.7 $\\ 
GX~339--4 &$( 338.94 , -4.33 )$&$ 10 _{ -4 }^{ +5 }$ $^{k}$&$ 5.18 $\\ 
GX~340+0 &$( 339.59 , -0.08 )$&$ 11 $ $^{c}$&$ 20.0 $\\ 
GX~349+2 &$( 349.10 , 2.75 )$&$ 9.2 $ $^{c}$&$ 6.13 $\\ 
GX~354+0 &$( 354.30 , -0.15 )$&$ 5.3 \pm 0.8$ $^{e}$&$ 13.9 $\\ 
GX~5--1 &$( 5.08 , -1.02 )$&$ 0.21 \pm 0.01$ $^{l}$&$ 10.4 $\\ 
V4641~Sgr &$( 6.77 , -4.79 )$&$ -$&$ 3.23 $\\ %
X1543--62 &$( 321.76 , -6.34 )$&$ 7 $ $^{m}$&$ 3.79 $\\ 
X1822--371 &$( 356.85 , -11.29 )$&$ 2.5 \pm 0.5$ $^{n}$&$ 1.40 $\\ 
XTEJ1814--338 &$( 358.75 , -7.59 )$&$ -$&$ 2.29 $\\ %
4U~1735--44 &$( 346.05 , -6.99 )$&$ 9.4 \pm 1.4$ $^{e}$&$ 3.96 $\\ 
GX~9+1 &$( 9.08 , 1.15 )$&$ 4.4 \pm 1.3$ $^{o}$&$ 9.89 $\\ 
4U~1916--053 &$( 31.36 , -8.46 )$&$ 8.8 \pm 1.3$ $^{e}$&$ 3.72 $\\ 
4U~1957+11 &$( 51.31 , -9.33 )$&$ -$&$ 2.01 $\\ %
A1744-361 &$( 354.12 , -4.19 )$&$ 9  $ $^{p}$&$ 4.44 $\\ 
Cir~X--1&$( 322.12 , 0.04 )$&$ 9.2 _{ -1.4 }^{ +1.3 }$ $^{e}$&$ 16.4 $\\ 
Cyg~X--2&$( 87.33 , -11.32 )$&$ 13.4 _{ -2 }^{ +1.9 }$ $^{e}$&$ 3.09 $\\ 
Ser~X--1&$( 36.12 , 4.84 )$&$ 11.1 \pm 1.6$ $^{e}$&$ 5.42 $\\ 
\hline
\multicolumn{4}{l}{ $N({\rm HI})$ in units of $10^{21}$cm$^{-2}$ }\\
\multicolumn{4}{l}{Distances obtained from$^a$\citet{pae01b}; }\\
\multicolumn{4}{l}{$^b$\citet{int03}; $^c$\citet{gri02}; }\\
\multicolumn{4}{l}{$^d$\citet{gal06};$^e$\citet{jon04};}\\
\multicolumn{4}{l}{$^f$\citet{cor05};$^g$\citet{bau18};}\\ 
\multicolumn{4}{l}{$^h$\citet{kon00};$^i$\citet{ban99};}\\ 
\multicolumn{4}{l}{$^j$\citet{oos01};$^k$\citet{hyn04};}\\ 
\multicolumn{4}{l}{$^l$\citet{gai20};$^m$\citet{wan04};}\\ 
\multicolumn{4}{l}{$^n$\citet{mas82};$^o$\citet{iar05};}\\ 
\multicolumn{4}{l}{$^p$\citet{bha06}.} 
\end{tabular} 
\end{table}

             \begin{figure*}
          \centering
\includegraphics[width=0.33\textwidth]{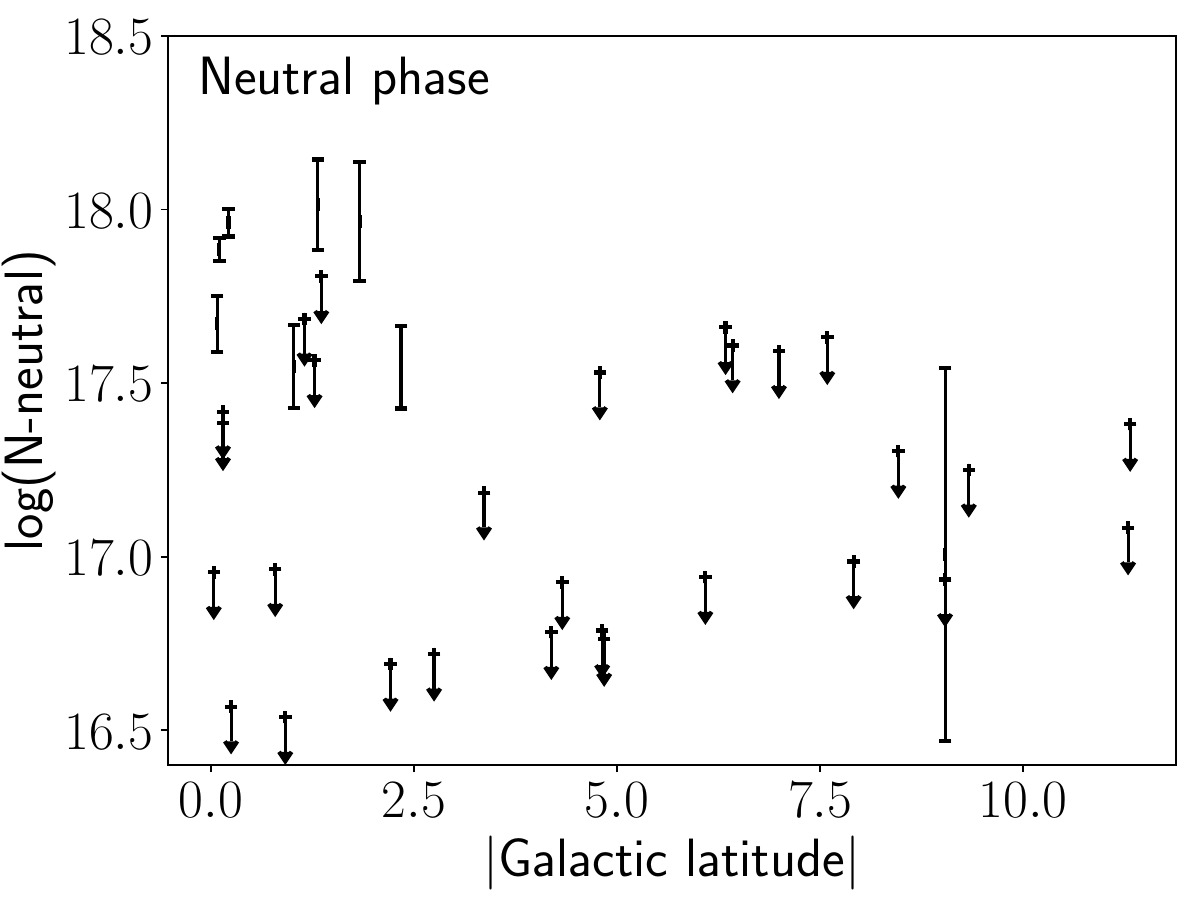} 
\includegraphics[width=0.33\textwidth]{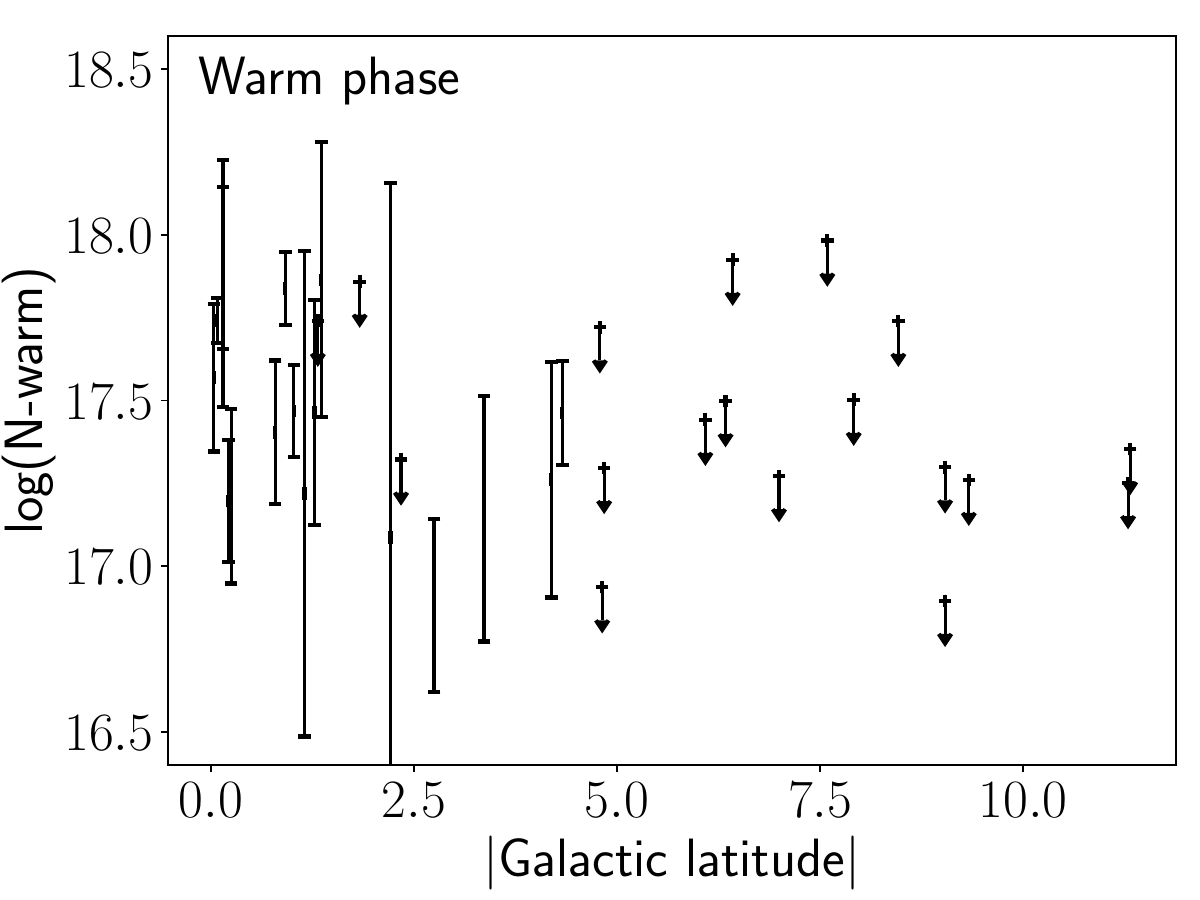} 
\includegraphics[width=0.33\textwidth]{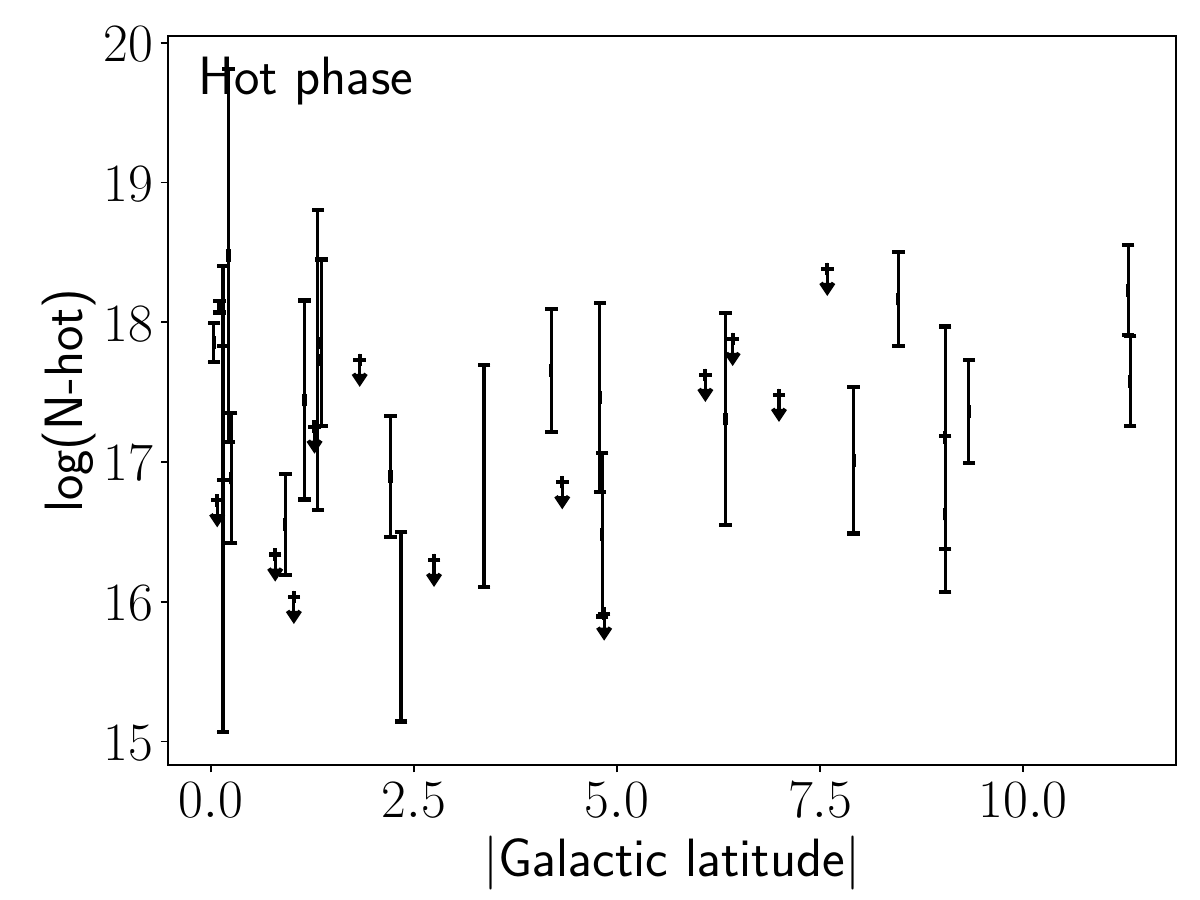} \\
\includegraphics[width=0.33\textwidth]{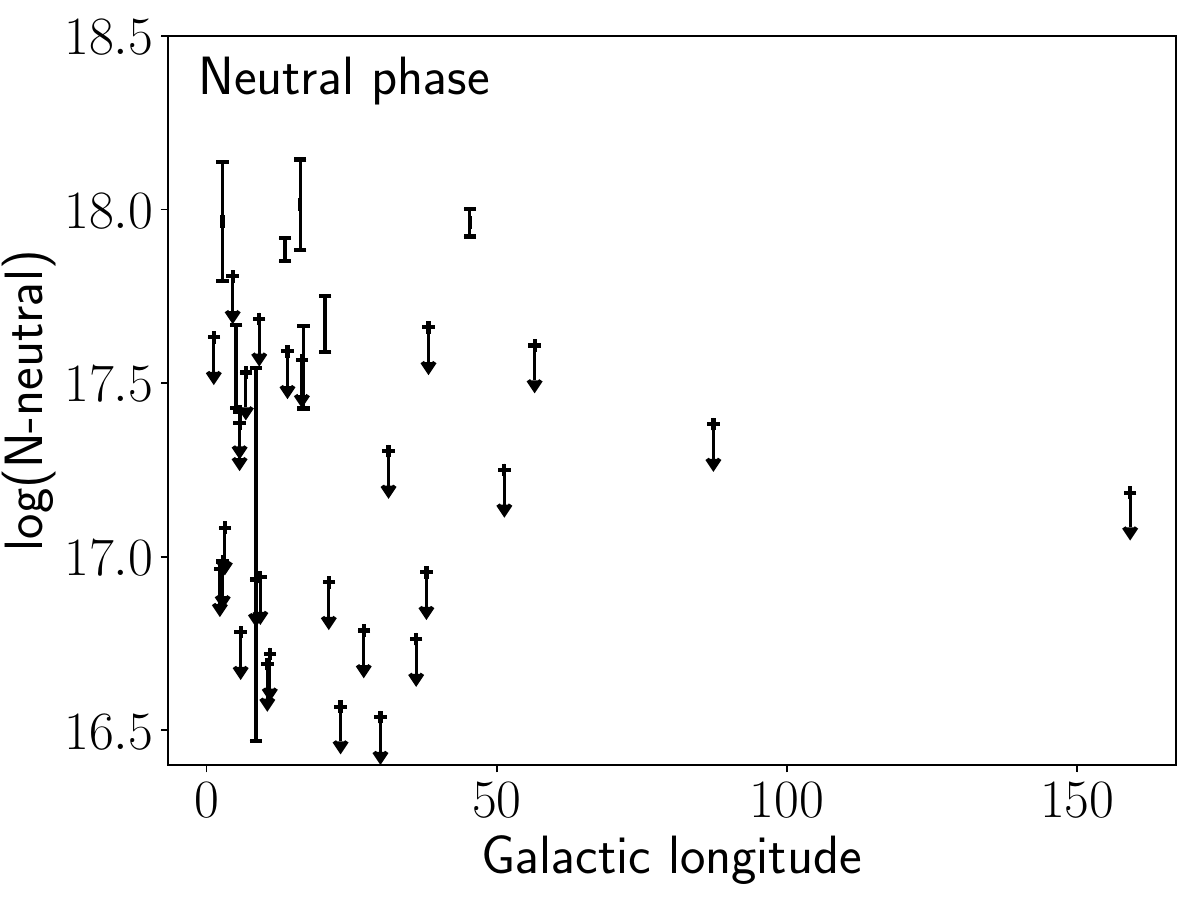} 
\includegraphics[width=0.33\textwidth]{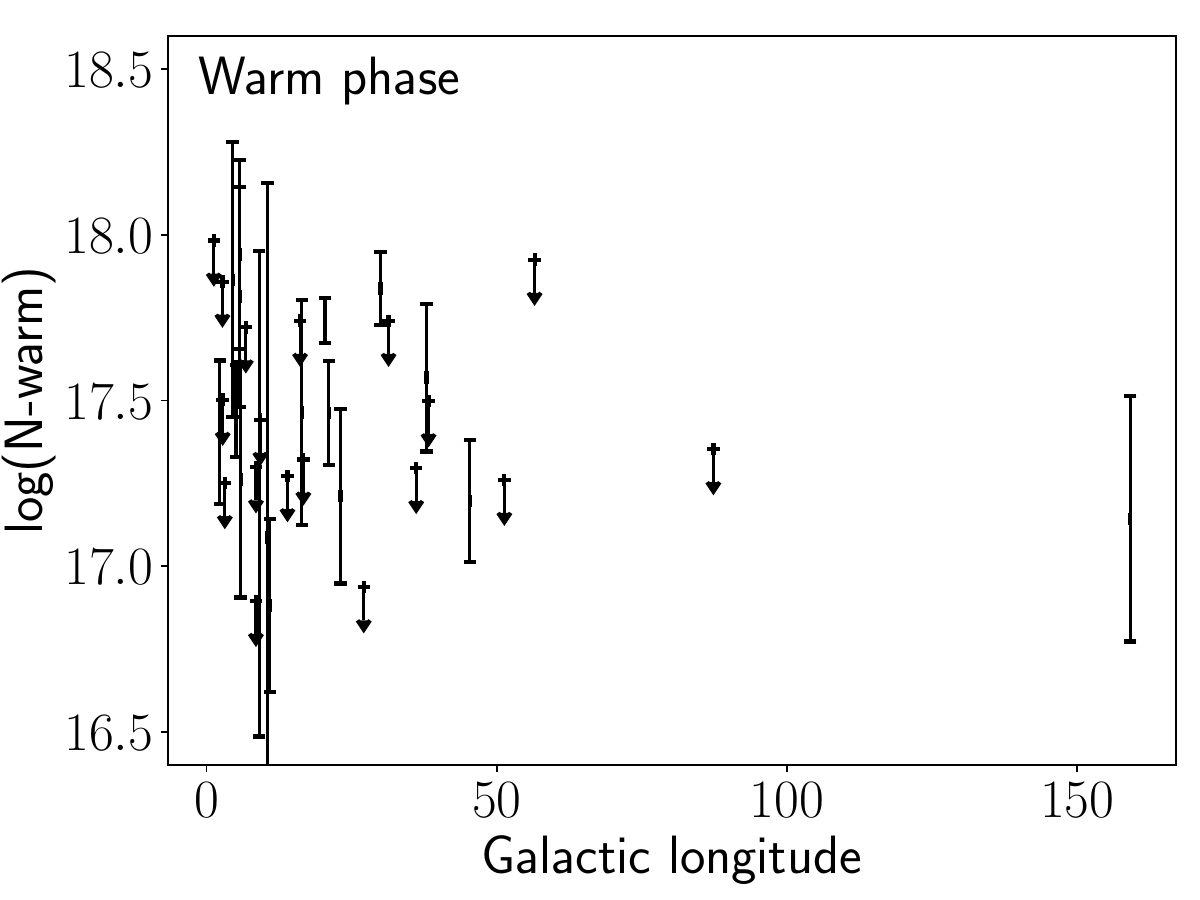} 
\includegraphics[width=0.33\textwidth]{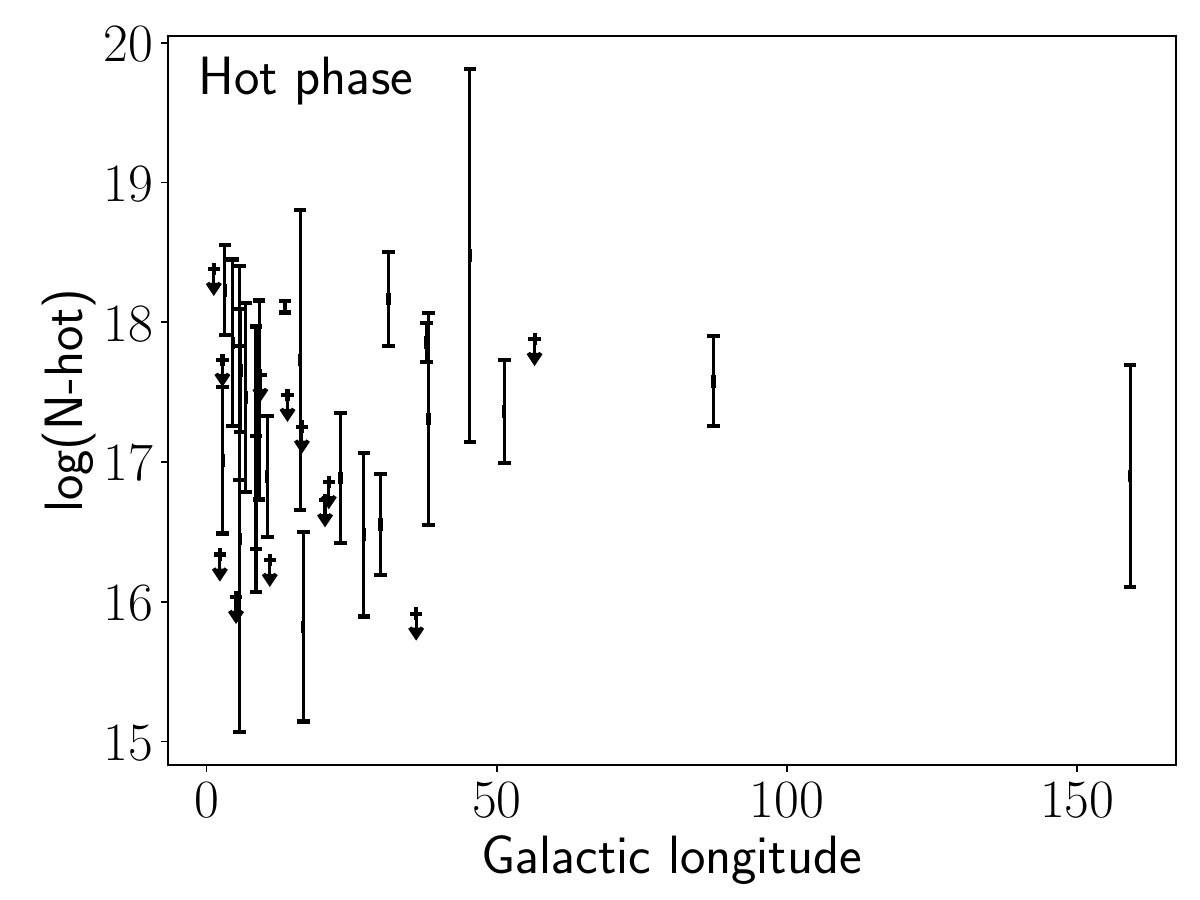} \\ 
      \caption{
      S column densities distribution for each ISM phase as function of Galactic latitude (top panels) and Galactic longitude (bottom panels).
      }\label{fig_columns_latitude}
   \end{figure*}

             \begin{figure}
          \centering
\includegraphics[width=0.48\textwidth]{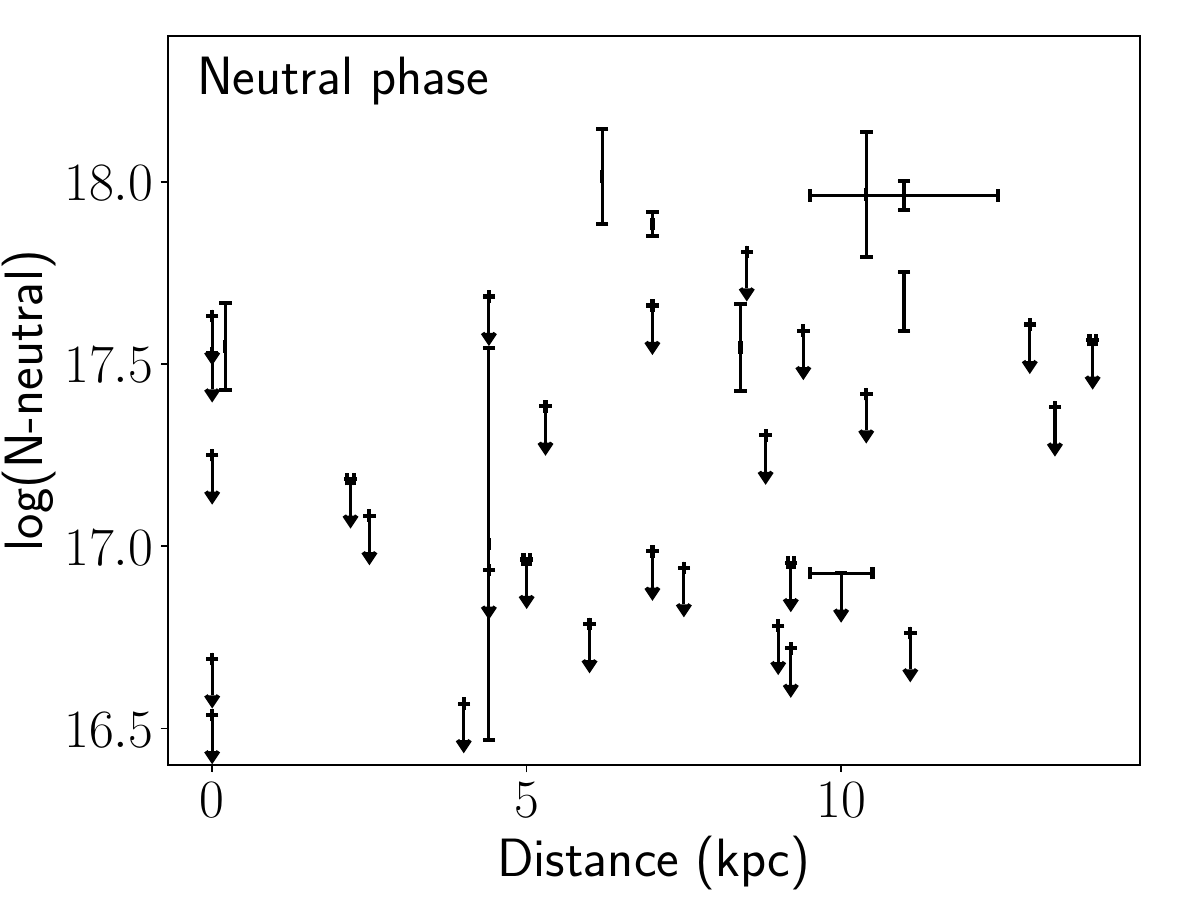} \\
\includegraphics[width=0.48\textwidth]{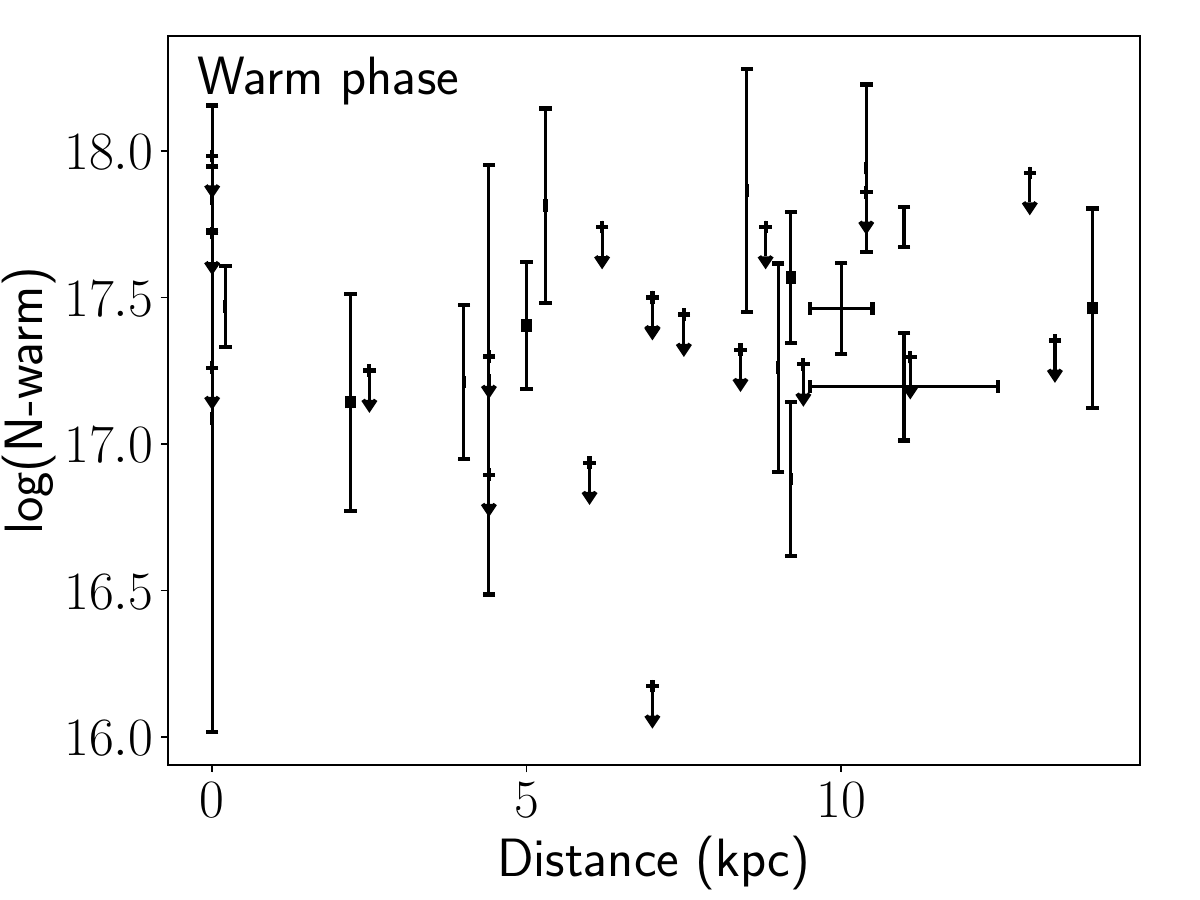} \\
\includegraphics[width=0.48\textwidth]{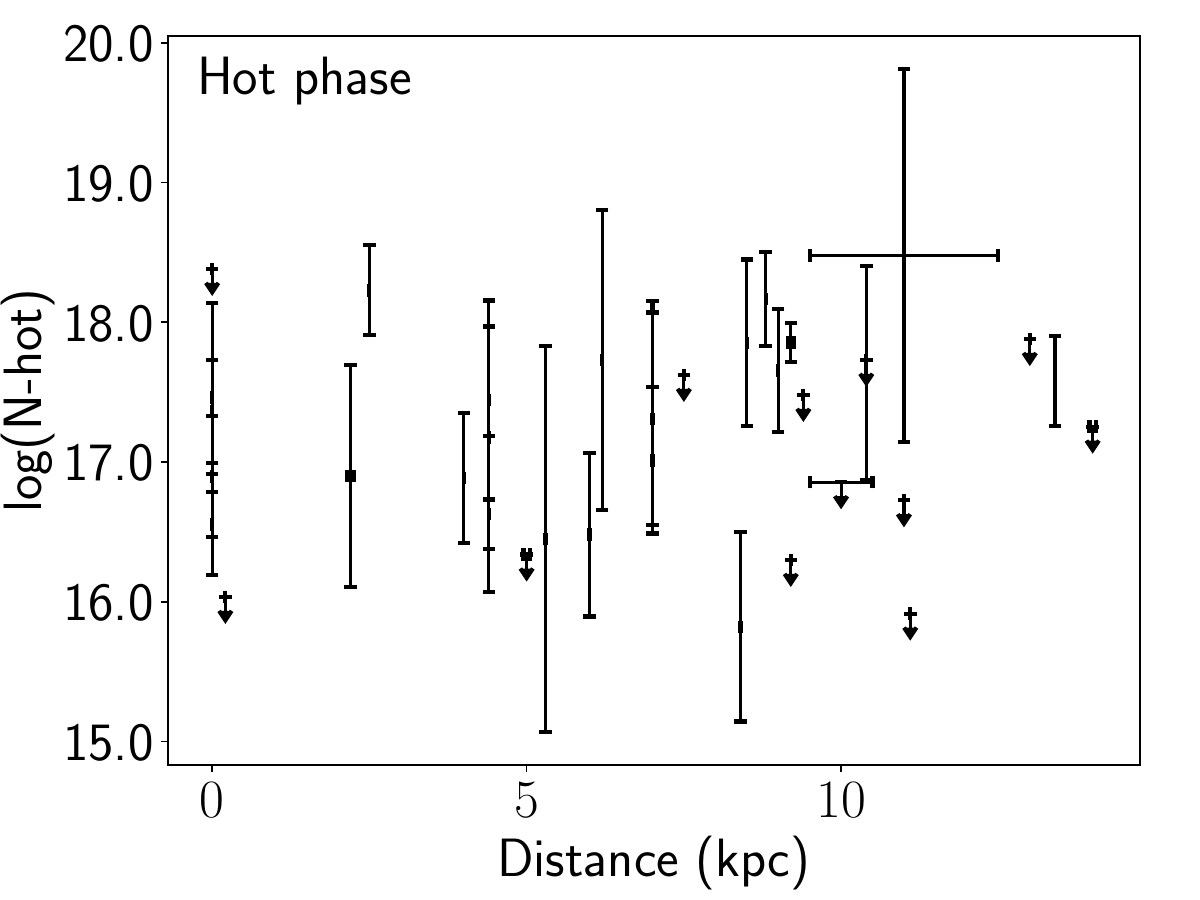} 
      \caption{
      S column densities distribution for each ISM phase as function of the distance.
      }\label{fig_columns_distance}
   \end{figure}

\section{Results and discussion}\label{sec_dis}
Table~\ref{tab_ismabs} shows the best-fit results. We have found good fits to the observed sulfur K-edge spectra for all observations. We noted that for most of the sources, we had obtained upper limits for the relevant parameters, including the dust component. We identified the different phases of the gaseous ISM as cold ({\rm S}~{\sc i}), warm ({\rm S}~{\sc ii}+{\rm S}~{\sc iii}), hot ({\rm S}~{\sc xiv}+{\rm S}~{\sc xv}+{\rm S}~{\sc xvi}) and dust (alabandite+pyrrohtite+troilite). The best-fit column densities obtained are shown in Figure~\ref{fig_data_density_fractions}. We note that the column densities between the different gaseous phases are in the same range, while the dust column densities tend to be much lower than the cold component.

Figure~\ref{fig_columns_latitude} shows the column density distribution for each ISM gaseous phase as a function of the Galactic latitude (top panels) and the Galactic longitude (bottom panels). The cold-warm phases tend to decrease with the Galactic latitude (top panels), while the hot phase does not appear to show a clear correlation, although we note that, for many sources, the best-fit results correspond to upper limits. \citet{gat21} found a similar homogeneous distribution for the warm-hot ISM component in their nitrogen K-edge photoabsorption region analysis. It is commonly assumed that the neutral component of the ISM exponentially decreases along the perpendicular direction to the Galactic plane, with larger column densities near the Galactic center \citep[see, e.g.][]{rob03,kalb09}. However, the sulfur depletion into dust may lead to departures from such distribution. Figure~\ref{fig_columns_distance} shows the column density distribution for each phase as a function of the distance for those sources where available. There is no clear correlation between both parameters.

Previous analysis of the ISM using X-ray absorption have shown that the gaseous state is dominated by the neutral component with mass fractions for the different phases in the Galactic disc of $\sim~ 90\%$, warm $\sim  8\%$ and hot $\sim  2\%$ of the total contribution \citep[e.g.][]{yao06,pin13,gat18a}. Given the uncertainties in the obtained column densities, we cannot accurately compute mass fractions for all sources. Moreover, we are not considering ionization equilibrium for the sulfur ionic species because the column densities in the {\it ISMabs} model are free parameters. Therefore, the hot phase temperature, as described in previous works, could not be hot enough in order to produce highly ionized S. Furthermore, the hot phase could include a contribution from an ionized static absorber intrinsic to the source \citep[see, for example][]{gat20b}. A complete thermodynamic analysis of the ISM component is beyond the scope of this work, which focuses on measuring the column densities for the neutral and ionic S species. For the dust component, we have found upper limits for all sources, with a contribution of $< 10\%$ for the cold gas.

  \begin{table*}
\caption{
\label{tab_ismabs}
Best-fit sulfur column densities obtained.
 }
\scriptsize
\centering
\begin{tabular}{lcccccccccccccc}
\hline  
Source  & S\,{\sc i} &  S\,{\sc ii}  &  S\,{\sc iii} & S\,{\sc xiv} &  S\,{\sc xv}  &  S\,{\sc xvi} & Alabandite  & Pyrrohtite & Troilite   &$\chi^{2}$/d.of.\\
 \hline
\hline 
4U0614+091& $<15.3$& $<7.3$& $13.9\pm 8.5$& $4.3\pm 0.8$& $<2.4$& $3.3\pm 0.5$& $<8.1$& $<7.5$& $<7.8$& $1269.0/1259$\\
4U1254-690& $<40.6$& $<61.8$& $<22.4$& $<33.7$& $<12.8$& $<29.5$& $<26.9$& $<23.8$& $<25.5$& $355.9/308$\\
4U1630-472& $<3.7$& $<3.1$& $16.3\pm 9.0$& $<1.1$& $<2.9$& $7.3\pm 6.9$& $<12.4$& $<62.1$& $<104.4$& $1315.7/1259$\\
4U1636-53& $<6.1$& $<3.5$& $<5.2$& $<0.7$& $<1.3$& $2.8\pm 0.4$& $<4.6$& $<3.7$& $<4.4$& $3130.1/2846$\\
4U1702-429& $103.4\pm 31.4$& $<32.7$& $<22.4$& $<47.2$& $<13.8$& $<37.8$& $<0.1$& $<0.1$& $<19.0$& $292.8/308$\\
4U1705-44& $35.1\pm 10.3$& $<17.0$& $<4.0$& $<0.3$& $<0.9$& $<0.6$& $<2.3$& $<2.2$& $<2.1$& $1642.1/1576$\\
4U1728-16& $<8.6$& $<15.0$& $<4.9$& $14.3\pm 2.2$& $<4.4$& $<2.9$& $<0.1$& $<15.4$& $<0.1$& $322.5/308$\\
4U1728-34& $<26.2$& $84.7\pm 39.9$& $<24.3$& $<82.1$& $<26.5$& $<42.5$& $<22.6$& $<22.9$& $<21.9$& $334.7/308$\\
GX9+9& $10.2\pm 8.7$& $<5.0$& $<2.8$& $<4.6$& $<0.5$& $3.2\pm 2.1$& $<5.2$& $<5.0$& $<4.6$& $291.9/308$\\
H1743-322& $92.4\pm 35.4$& $<60.5$& $<11.7$& $<25.7$& $<25.2$& $<3.1$& $<0.1$& $<0.1$& $<0.1$& $2200.6/2527$\\
IGRJ17091-3624& $<4.9$& $<14.0$& $<6.8$& $<0.5$& $<1.1$& $7.8\pm 6.9$& $<4.5$& $<5.2$& $<4.3$& $2548.9/2527$\\
NGC6624& $<9.7$& $<6.4$& $<25.4$& $<6.2$& $<10.3$& $<1.0$& $<95.1$& $<42.9$& $<35.4$& $612.3/625$\\
EXO1846-031& $<3.5$& $35.8\pm 9.8$& $33.1\pm 7.2$& $<0.2$& $<0.1$& $3.6\pm 2.9$& $<10.8$& $<10.8$& $<9.3$& $1660.7/1576$\\
GRS1758-258& $<64.2$& $<70.5$& $73.2\pm 41.1$& $<71.3$& $<4.6$& $<4.7$& $<39.2$& $<22.5$& $<28.2$& $637.8/625$\\
GRS1915+105& $91.7\pm 8.5$& $<1.2$& $15.7\pm 6.0$& $<0.3$& $<0.3$& $30\pm 14.6$& $<12.0$& $<11.0$& $<10.0$& $1809.0/1579$\\
GS1826-238& $<8.7$& $<9.8$& $<17.9$& $<32.5$& $<1.2$& $<8.3$& $<9.9$& $<9.3$& $<8.7$& $329.9/308$\\
GX13+1& $76.7\pm 5.8$& $<0.7$& $<0.8$& $<0.6$& $<0.1$& $129.7\pm 12.5$& $<2.1$& $<2.1$& $<2.3$& $3182.2/2844$\\
GX17+2& $<36.8$& $<30.1$& $24.1\pm 10.8$& $<13.8$& $<0.7$& $<3.4$& $<4.6$& $<0.1$& $<0.1$& $281.8/308$\\
GX3+1& $<9.2$& $22.6\pm 7.9$& $<7.2$& $<1.2$& $<0.1$& $<0.8$& $<10.8$& $<10.4$& $<10.0$& $3099.2/3161$\\
GX339-4& $<8.4$& $<4.6$& $29.0\pm 8.3$& $<6.5$& $<0.4$& $<0.3$& $<19.4$& $<8.1$& $<8.6$& $953.5/942$\\
GX340+0& $46.9\pm 8.6$& $<3.8$& $55.1\pm 6.8$& $<4.8$& $<0.3$& $<0.2$& $<20.0$& $<18.1$& $<18.0$& $1341.3/1262$\\
GX349+2& $<5.3$& $<5.9$& $7.6\pm 3.1$& $<1.5$& $<0.1$& $<0.4$& $<7.1$& $<7.1$& $<7.3$& $2069.7/1893$\\
GX354+0& $<24.2$& $38.8\pm 24.3$& $26.3\pm 16.7$& $<1.7$& $2.2\pm 0.44$& $<6.7$& $<7.3$& $<0.1$& $<7.1$& $821.9/942$\\
GX5-1& $35.3\pm 9.2$& $22.9\pm 6.8$& $6.5\pm 3.1$& $<0.9$& $<0.0$& $<0.1$& $<0.1$& $<1.4$& $<0.1$& $2558.3/2527$\\
V4641Sgr& $<33.9$& $<42.4$& $<10.1$& $<16.5$& $<1.3$& $28.8\pm 3.62$& $<20.0$& $<0.1$& $<20.3$& $565.3/625$\\
X1543-62& $<45.8$& $<17.9$& $<13.6$& $<7.1$& $20.2\pm 2.9$& $<9.6$& $<64.0$& $<55.7$& $<58.4$& $306.9/308$\\
X1822-371& $<12.1$& $<10.3$& $<7.6$& $<16.8$& $<11.8$& $<0.7$& $<0.1$& $<9.9$& $<0.1$& $533.4/625$\\
XTEJ1814-338& $<42.9$& $<43.9$& $<52.2$& $<0.0$& $<181.0$& $<60.1$& $<49.8$& $<48.8$& $<48.9$& $317.1/308$\\
4U1735-44& $<39.1$& $<11.4$& $<7.3$& $<25.1$& $<4.4$& $<0.8$& $<16.8$& $<20.6$& $<19.4$& $346.4/308$\\
GX9+1& $<48.5$& $<35.6$& $16.3\pm 12.3$& $27.8\pm 19.9$& $<1.2$& $<4.4$& $<44.4$& $<44.0$& $<43.2$& $637.5/625$\\
4U1916-053& $<20.2$& $<40.1$& $<14.9$& $57.9\pm 33.7$& $<4.6$& $88.5\pm 48.2$& $<31.8$& $<39.9$& $<57.0$& $1966.7/1893$\\
4U1957+11& $<17.8$& $<8.5$& $<9.7$& $<1.2$& $<0.6$& $23.0\pm 18.0$& $<0.1$& $<0.1$& $<9.6$& $946.3/942$\\
A1744-361& $<6.1$& $<5.7$& $18.2\pm 11.4$& $43.1\pm 36.8$& $<1.3$& $<11.2$& $<4.9$& $<4.6$& $<4.8$& $884.7/942$\\
CIRX-1& $<9.0$& $25.7\pm 10.9$& $11.3\pm 8.7$& $<0.7$& $<4.1$& $71.1\pm 21.0$& $<7.0$& $<6.7$& $<3.7$& $1366.0/1259$\\
CYGX-2& $<24.1$& $<11.7$& $<10.9$& $35.0\pm 22.9$& $1.9\pm 1.3$& $<4.3$& $<28.7$& $<30.2$& $<30.3$& $740.0/625$\\
SERX-1& $<5.8$& $<13.3$& $<6.5$& $<0.2$& $<0.5$& $<0.1$& $<0.1$& $<3.4$& $<0.1$& $978.7/942$\\
\hline 
\multicolumn{10}{l}{ Column densities in units of $10^{16}$cm$^{-2}$ for the ionic species and $10^{14}$cm$^{-2}$ for the dust samples.}
 \end{tabular}
\end{table*}

  \subsection{Future prospects}\label{sec_sim}
\begin{figure}
          \centering
\includegraphics[width=0.48\textwidth]{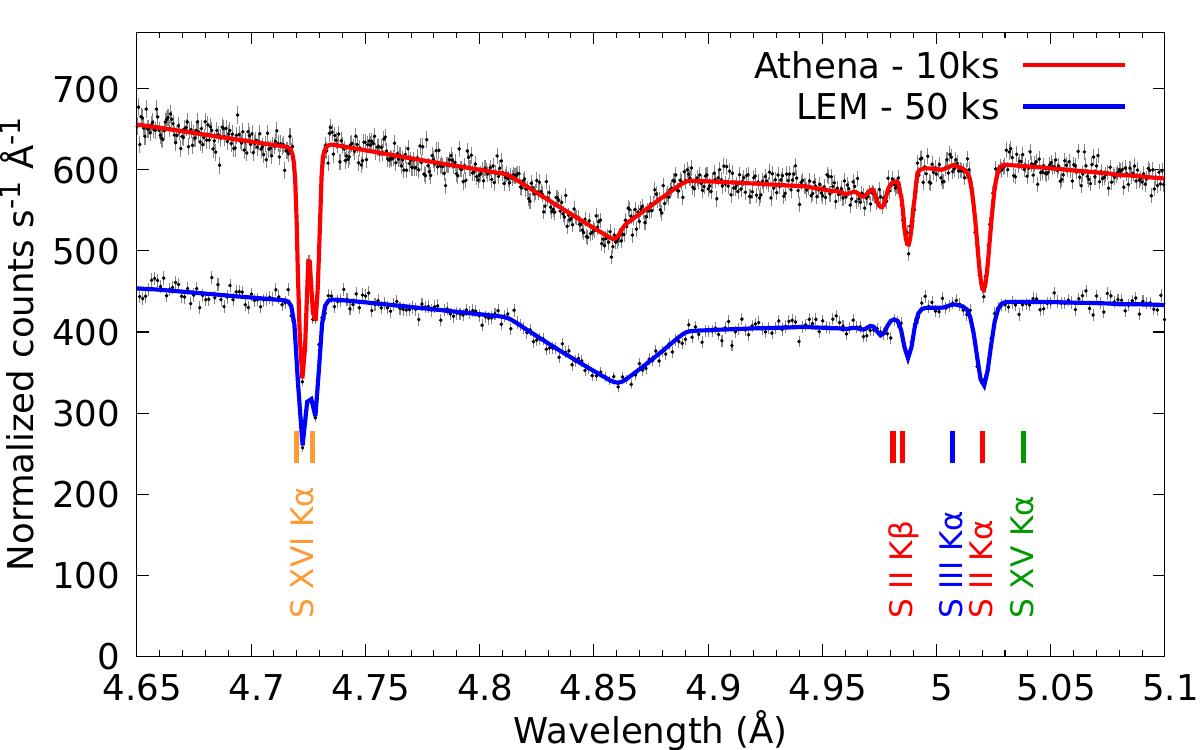} 
      \caption{
      {\it Athena} and {\it LEM} simulation of the S K-edge photoabsorption region for a Galactic source (e.g. GX13+1). The total exposure time is indicated. 
      }\label{ath_sim}
   \end{figure}

Future X-ray observatories will allow us to resolve the $K\alpha$ resonance lines for the different S ionic species. For the neutral {\rm S}~{\sc i} component, it is crucial to compute new accurate atomic data, including radiation and Auger damping, in order to model not only the gaseous phase but also to differentiate it with the modulations and spectral features induced by interstellar grains extinction \citep{cos19b}. For example, Figure~\ref{ath_sim} shows simulations for a Galactic source (i.e. GX13+1) obtained with {\it Athena} \citep{nan13} and {\it LEM} \citep{kra22}. The plot shows the extraordinary capabilities of the different instruments, where the main resonance absorption lines are easily visible, including the {\rm S}~{\sc i} $K$ resonance lines. Moreover, by simultaneously measuring K$\alpha$ and K$\beta$ absorption lines for the same ions, more accurate constraints on the abundances, broadening, and ionization state of the ISM absorber will be obtained. It is important to note that this simulation includes only the gaseous component. While the total dust contribution could also be measured, the distinction between different dust samples is more challenging. A detailed dust simulation for {\it Athena} was computed by \citet{cos19b}.

\section{Conclusions}\label{sec_con}
We have analyzed the sulfur K-edge X-ray absorption region (4.5-5.5 \AA) using {\it Chandra} high-resolution spectra of 36 LMXBs. We fitted each source with a simple {\tt powerlaw} for the continuum and a modified version of the {\it ISMabs} model, including extinction cross-sections for three samples of interstellar dust analogs (e.g., alabandite, pyrrhotite, and troilite). We found that the absorption features identified in the spectra are well modeled with the theoretical photoabsorption cross-section, even though individual K$\alpha$ doublets/triplets cannot be resolved. With this model, we have estimated column densities for {\rm S}~{\sc i}, {\rm S}~{\sc ii}, {\rm S}~{\sc iii}, {\rm S}~{\sc v}, {\rm S}~{\sc vi} and {\rm S}~{\sc vii} ionic species as well as upper limits for the dust component, which trace the multiphase ISM. While the cold-warm column densities tend to decrease with the Galactic latitude, we found no correlation with distances or Galactic longitude. Finally, our simulations using response files from future X-ray observatories such as {\it Athena} indicate that a detailed benchmarking of the atomic data will be possible with such instruments.

\subsection*{Data availability}
Observations analyzed in this article are available in the Chandra Grating-Data Archive and Catalog (TGCat)  (\url{http://tgcat.mit.edu/about.php}). The {\tt ISMabs} model is included in the {\sc xspec} data analysis software (\url{https://heasarc.gsfc.nasa.gov/xanadu/xspec/}). This research was carried out on the High Performance Computing resources of the cobra cluster at the Max Planck Comput-ing and Data Facility (MPCDF) in Garching operated by the Max Planck Society (MPG)

 \subsection*{Acknowledgements}
We thank the anonymous referee for the careful reading of our manuscript and the valuable comments.

\bibliographystyle{mnras}

\end{document}